# THE NON-SINGLET QCD EVOLUTION KERNELS IMPROVED BY RENORMALON CHAIN CONTRIBUTIONS

S. V. Mikhailov [1]

*Joint Institute for Nuclear Research, Bogoliubov Laboratory of Theoretical Physics,*
*141980, Moscow Region, Dubna, Russia*

**Abstract**

Closed expressions are presented for the contributions to QCD non-singlet forward evolution kernels $P(z)$ for the DGLAP equation and to $V(x, y)$ for non-forward (ER-BL) evolution equation for a certain class of diagrams which include renormalon chains. Calculations are performed in covariant $\xi$-gauge, in the $\overline{\mathrm{MS}}$ scheme. The assumption of "naive nonabelianization" approximation for kernel calculations is discussed, and a special choice of the gauge parameter $\xi = -3$ is analyzed in this context. Partial solutions to the ER-BL evolution equation are obtained.

---

[1] E-mail: mikhs@thsun1.jinr.dubna.su

# 1 Introduction

Evolution kernels are the main ingredients of the well-known evolution equations for parton distribution of DIS processes and for parton wave functions in hard exclusive reactions. These equations describe the dependence of parton distribution functions and parton wave functions on the renormalization parameter $\mu^2$. Here I discuss the diagrammatic analysis and multiloop calculation of the forward DGLAP evolution kernel $P(z)$ [1] and non-forward Efremov-Radyushkin–Brodsky-Lepage (ER-BL) kernel $V(x,y)$ [2] in a class of "all-order" approximation of the perturbative QCD. The regular method of calculation and resummation of certain classes of diagrams for these kernels has been suggested in [3]. These diagrams include the chains of one-loop self-energy parts (renormalon chains) into the one-loop diagrams (see Fig. 1). Here the results for both the kinds of kernels (DGLAP and ER-BL), obtained earlier in the framework of a scalar model in six dimensions with the Lagrangian $L_{int} = g \sum_i^{N_f} (\psi_i^* \psi_i \varphi)_{(6)}$ with the scalar "quark" flavours ($\psi_i$) and "gluon" ($\varphi$), are extended to the non-singlet QCD kernels. For the readers convenience some important results of the paper [3] would be reminded.

The insertion of the chain into "gluon" line ("chain-1" in [3]) of the diagram in Fig.1 a,b and resummation over all bubbles lead to the transformation of the one-loop kernel ( see, *e.g.*, [4]) $aP_0(z) = a\bar{z} \equiv a(1-z)$ into the kernel $P^{(1)}(z;A)$

$$aP_0(z) = a\bar{z} \stackrel{chain-1}{\longrightarrow} P^{(1)}(z;A) = a\bar{z}\left[(z)^{-A}(1-A)\frac{\gamma_\varphi(0)}{\gamma_\varphi(A)}\right]; \text{ where } A = aN_f\gamma_\varphi(0), \ a = \frac{g^2}{(4\pi)^3}. \quad (1)$$

Here, $\gamma_{\psi(\varphi)}(\varepsilon)$ are the one-loop coefficients of the anomalous dimensions of quark (gluon at $N_f = 1$) fields in D-dimension ($D = 6-2\varepsilon$) discussed in [3]; for the scalar model $\gamma_\psi(\varepsilon) = \gamma_\varphi(\varepsilon) = B(2-\varepsilon, 2-\varepsilon)C(\varepsilon)$, and $C(\varepsilon)$ is a scheme-dependent factor corresponding to a certain choice of an $\overline{MS}$ –like scheme. The argument $A$ of the function $\gamma_\varphi(A)$ in (1) is the standard anomalous dimension (AD) of a gluon field. So, one can conclude that the "all-order" result in (1) is completely determined by the single quark bubble diagram. The resummation of this "chain-1" subseries into an analytic function in $A$ shouldn't be taken by surprise. Really, the considered problem can be connected with the calculation of large $N_f$ asymptotics of the AD's in order of $1/N_f$. An approach was suggested by A. Vasil'ev and collaborators at the beginning of 80' [5] to calculate the renormalization-group functions in this limit, they used the conformal properties of the theory at the critical point $g = g_c$ corresponding to the non-trivial zero $g_c$ of the D-dimensional $\beta$-function. This approach has been extended by J. Gracey for calculation of the AD's of the composite operators of DIS in QCD in any order $n$ of PT, [6]. I have used another approach, which is close to [7]; contrary to the large $N_f$ asymptotic method, it does not appeal to the value of parameters $N_f T_R$, $C_A/2$ or $C_F$, associated in QCD with different kinds of loops. To illustrate this feature, let us consider the insertions of chains of one-loop self-energy parts into the "quark" line of diagram Fig.1a ("chain-2" in [3]). Contributions of these diagrams, calculated in the framework of the above scalar model, do not contain the parameter $N_f$ and can be summarized into the kernel $P^{(2)}(z;B)$ [3]

$$aP_0(z) = a\bar{z} \stackrel{chain-2}{\longrightarrow} P^{(2)}(z;B) = a\bar{z}\left(1 + B\frac{d}{dB}\right)\left[(\bar{z})^{-B}\frac{\gamma_\psi(0)}{\gamma_\psi(B)}\right]; \text{ where } B = a\gamma_\psi(0), \quad (2)$$

according to the same approach. Following this way, the "improved" QCD kernel $P^{(1)}(z;A)$ has been obtained in [3] for the case of quark or gluon bubble chain insertions in the Feynman gauge and in [8] for general case of the mixed insertions in $\xi$– gauge.

In this talk, we present the QCD results similar to Eq.(1), for each type of diagrams appearing in the covariant $\xi$– gauge for the DGLAP non-singlet kernel $P(z; A)$. The analytic properties of the function $P(z; A)$ in variable $A$ are analyzed. The assumption of "Naive Nonabelianization" (NNA) approximation [9] for the kernel calculation [10] is discussed and its deficiency is demonstrated. The ER-BL evolution kernel $V(x, y)$ is obtained in the same approximation as the DGLAP kernel, by using the exact relations between $P$ and $V$ kernels [11, 3] for a class of "triangular diagrams" in Fig. 1. The considered class of diagrams represents the leading $N_f$ contributions to both kinds of kernels. The partial solutions for the ER-BL equation are derived (compare with [10, 12]).

## 2 Triangular diagrams for the DGLAP evolution kernel

Here, the results of the bubble chain resummation for QCD diagrams in Fig.1 a,b,c for the DGLAP kernel are discussed. These classes of diagrams generate contributions $\sim a_s (a_s \ln[1/z])^n$ in any order $n$ of PT. Based on the resummation method of Ref. [3] in the QCD version, one can derive the kernels $P^{(1a,b,c)}$ (corresponding to the diagrams in Fig.1) in the covariant $\xi$−gauge[2]

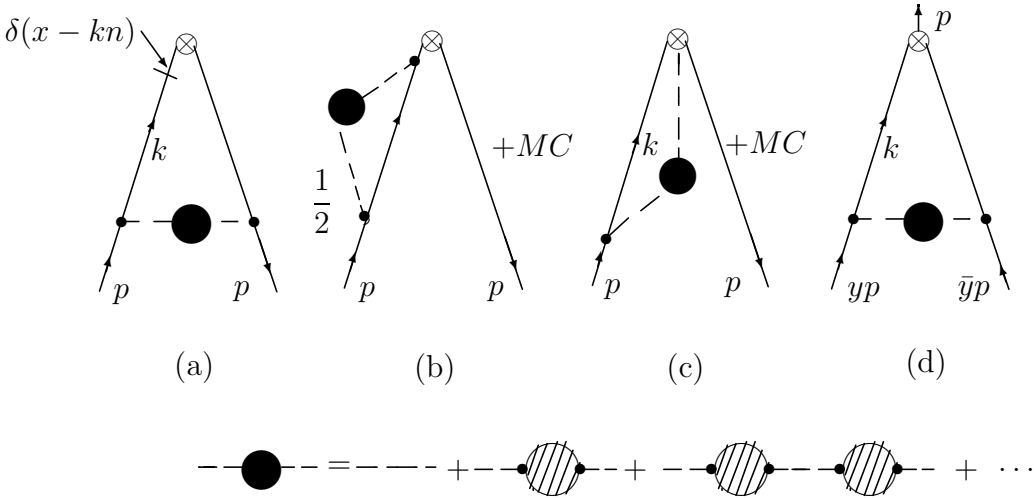

Figure 1: The diagrams in figs. 1a – 1c are the "triangular" diagrams for the QCD DGLAP kernel; dashed line for gluons, solid line for quarks; black circle denotes the sum of all kinds of the one-loop insertions (dashed circle), both quark and gluon (ghost) or mixed chains; MC denotes the mirror–conjugate diagram; 1d is an example of diagram for the non-forward ER-BL kernel.

$$P^{(1a)}(z; A) = a_s C_F 2\bar{z} \cdot (1 - A)^2 z^{-A} \frac{\gamma_g(0)}{\gamma_g(A)} - a_s C_F \cdot \delta(1 - z) \left( \frac{1}{(1 - A)} \frac{\gamma_g(0)}{\gamma_g(A)} - \xi \right), \quad (3)$$

$$P^{(1b)}(z; A) = a_s C_F 2 \cdot \left( \frac{2z^{1-A}}{1 - z} \frac{\gamma_g(0)}{\gamma_g(A)} \right)_+, \quad (4)$$

$$P^{(1c)}(z; A) = a_s C_F \cdot \delta(1 - z) \left( \frac{A(3 - 2A)}{(2 - A)(1 - A)} \frac{\gamma_g(0)}{\gamma_g(A)} - \xi \right), \quad (5)$$

where $a_s = \dfrac{\alpha_s}{4\pi}$, $C_F = (N_c^2 - 1)/2N_c$, $C_A = N_c$ and $T_R = \dfrac{1}{2}$ are the Casimirs of SU($N_c$) group, and $A = -a_s \gamma_g(0)$. The function $\gamma_g(\varepsilon)$ is the one-loop coefficient of the anomalous dimension of

---
[2]The gauge parameter $\xi$ is defined via the gluon propagator in lowest order $iD_{\mu\nu}(k^2) = \dfrac{-i\delta^{ab}}{k^2 + i\epsilon} \left( g_{\mu\nu} + (\xi - 1) \dfrac{k_\mu k_\mu}{k^2} \right)$

gluon field in D-dimension, here $D = 4 - 2\varepsilon$. In other words, it is the coefficient $Z_1(\varepsilon)$ of a simple pole in the expansion of the gluon field renormalization constant $Z$, that includes both its finite part and all the powers of the $\varepsilon$-expansion. The function $\gamma_g(\varepsilon)$ thus defined is an analytic function in the variable $\varepsilon$ by construction. Equations (3) – (5) are valid for any kind of insertions, i.e., $\gamma_g = \gamma_g^{(q)}$ for the quark loop, $\gamma_g = \gamma_g^{(g)}$ for the gluon (ghost) loop, or for their sum

$$\gamma_g(A, \xi) = \gamma_g^{(q)}(A) + \gamma_g^{(g)}(A, \xi);$$

when both kinds of insertions are taken into account. The sum of contributions (3), (4), (5) results in $P^{(1)}(z; A, \xi)$ which has the expected "plus form"

$$P^{(1)}(z; A, \xi) = a_s C_F 2 \cdot \left[\bar{z} z^{-A}(1-A)^2 + \frac{2z^{1-A}}{1-z}\right]_+ \frac{\gamma_g(0, \xi)}{\gamma_g(A, \xi)}, \tag{6}$$

$$a_s P_0(z) = a_s C_F 2 \cdot \left[\bar{z} + \frac{2z}{1-z}\right]_+, \tag{7}$$

here, for comparison with (6), the one-loop result $a_s P_0(z)$ is written down, the latter can be obtained as the limit $P^{(1)}(z; A \to 0, \xi)$. Note that in (6) the $\delta(1-z)$ - terms are exactly accumulated in the form of the $[\ldots]_+$ prescription, and the $\xi$ - terms successfully cancel. This is due to the evident current conservation for the case of quark bubble insertions, including the gluon bubbles into consideration merely modifies the effective AD, $\gamma_g(A, \xi) \to \gamma_g^{(q)}(A)$, conserving the structure of the result (6), see [3, 8]. Substituting the well-known expressions for $\gamma_g(\varepsilon)$ from the quark or gluon (ghost) loops (see, e.g., [13])

$$\gamma_g^{(q)}(\varepsilon) = -8 N_f T_R B(D/2, D/2) C(\varepsilon), \tag{8}$$

$$\gamma_g^{(g)}(\varepsilon, \xi) = \frac{C_A}{2} B(D/2 - 1, D/2 - 1) \left(\left(\frac{3D-2}{D-1}\right) + \right.$$

$$\left. (1-\xi)(D-3) + \left(\frac{1-\xi}{2}\right)^2 \varepsilon \right) C(\varepsilon), \tag{9}$$

into the general formulae (3) – (5), and (6) one can obtain $P^{(1)}(z; A, \xi)$ for both the quark and the gluon loop insertions simultaneously. Here, the coefficient $C(\varepsilon) = \Gamma(1-\varepsilon)\Gamma(1+\varepsilon)$ implies a certain choice of the $\overline{\text{MS}}$ scheme where every loop integral is multiplied by the scheme factor $\Gamma(D/2 - 1)(\mu^2/4\pi)^\varepsilon$. The renormalization scheme dependence of $P^{(1)}(z; A)$ is accumulated by the factor $C(\varepsilon)$ [3]. Of course, the final result (6) will be gauge-dependent in virtue of the evident gauge dependence of the gluon loop contribution $\gamma_g^{(g)}(\varepsilon, \xi)$, in this case, e.g.,

$$A(\xi) = -a_s \gamma_g(0, \xi) = -a_s \left(\gamma_g^{(g)}(0, \xi) + \gamma_g^{(q)}(0)\right) = -a_s \left[\left(\frac{5}{3} + \frac{(1-\xi)}{2}\right) C_A - \frac{4}{3} N_f T_R\right], \tag{10}$$

is the contribution to the one-loop renormalization of the gluon field. The positions of zeros of the function $\gamma_g(A, \xi)$ in $A$, which manifest the poles of $P(z; A, \xi)$, also depend on $\xi$. The kernel $P^{(1)}(z; A, \xi)$ became gauge-invariant in the case when only the quark insertions are involved, i.e., $\gamma_g^{(q)} \to \gamma_g$; $A = A^{(q)} = -a_s \gamma_g^{(q)}(0) = a_s \frac{4}{3} T_R N_f$, and $P^{(1)}(z; A^{(q)}) \to P^{(1)}(z; A, \xi)$ as it was presented in [3]. It is instructive to consider this case in detail. To this end, let us choose the common factor $\gamma_g^{(q)}(0)/\gamma_g^{(q)}(A)$ in formula (6) for the crude measure of modification of the kernel in comparison with the one-loop result $a_s P_0(z)$. Considering the curve of this factor in the argument $A$ in Fig.2, one may conclude:

---

[3] For another popular definition of a minimal scheme, when a scheme factor is chosen as $\exp(c \cdot \varepsilon)$, $c = -\gamma_E + \ldots$ instead of $\Gamma(D/2 - 1)$, the coefficient $C(\varepsilon)$ does not contain any scheme "traces" in final expressions for the renormalization-group functions.

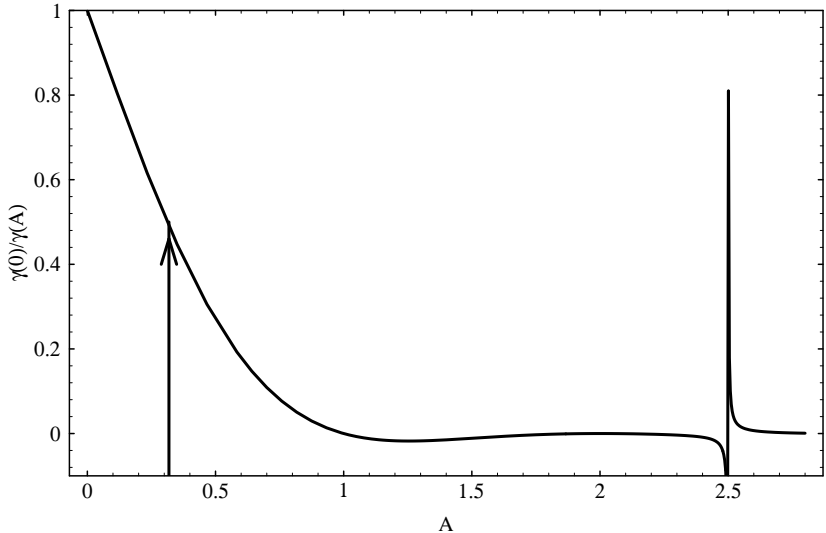

Figure 2: The curve of the factor $\gamma_g(0)/\gamma_g(A)$, the arrow on the picture corresponds to the point $A = 1/\pi$ (i.e. $\alpha_s = 6/N_f$), the first singularity appears at $A = A_0 = 5/2$.

(i) the range of convergence of PT series corresponds to the left zero of the function $\gamma_g^{(q)}(A)$ and is equal to $A_0 = 5/2$, that corresponds to $\alpha_s^0 = 15\pi/N_f$, so, this range looks very broad [4], $\alpha_s < 5\pi$ at $N_f = 3$;

(ii) in spite of a wide range of PT fidelity, the resummation into $P_q^{(1)}(z; A)$ is substantial – two zeros of the function $P_q^{(1)}(z; A)$ in $A$ appear within the range of convergence (it depends on a certain $\overline{MS}$ scheme);

(iii) the factor $\gamma_g^{(q)}(0)/\gamma_g^{(q)}(A)$ decays quickly with the growth of the argument $A$. Really, if we take the naive boundary of the standard PT applicability, $\alpha_s = 1$ at $N_f = 3$, $A^{(q)} = 1/(2\pi)$, then this factor falls approximately to 0.7, at $N_f = 6$, $A^{(q)} = 1/\pi$ it falls to 0.5, see arrow in Fig. 2; thus, the resummation is numerically important in this range.

Note at the end that Eq.(6) could not provide valid asymptotic behavior of the kernels for $z \to 0$. A similar $z$-behavior is determined by the double-logarithmic corrections which are most singular at zero, like $a_s \left( a_s \ln^2[z] \right)^n$ [14]. These contributions appear due to renormalization of the composite operator in the diagrams by ladder graphs, *etc.* rather than by the triangular ones.

## 3 Analysis of the NNA assumption for kernel calculations

The expansion of $P_q^{(1)}(z; A)$ in $A$ provides the leading $a_s (a_s N_f \ln[1/z])^n$ dependence of the kernels with a large number $N_f$ in any order $n$ of PT [3]. But these contributions do not numerically dominate for real numbers of flavours $N_f = 4, 5, 6$. That may be verified by comparing the total numerical results for the 2– and 3–loop AD's of composite operators (ADCO) presented in [16] with their $N_f$-leading terms (see ADCO in Table 1). Therefore, to obtain a satisfactory agreement at least with the second order results, one should take into account the contribution from **subleading** $N_f$-terms. As a first step, let us consider the contribution from the completed renormalization of the gluon line – it should generate **a part of subleading** terms. Below we shall examine an exceptional choice of the gauge parameter $\xi = -3$. For this gauge the coefficient of one-loop gluon AD $\gamma_g(0, -3)$ coincides with $b_0$, the one-loop coefficient of the $\beta$-function [5]. Therefore this gauge may be used for a reformulation of the so-called [9] NNA proposition to kernel calculations. Note, just this value of $\xi$ has been used in [15] to estimate the total gluon contribution only from the gluon bubble in order of $a_s^2$ to the process of $e^+ \, e^-$ annihilation.

---

[4]Here we consider the evolution kernel $P(z, A)$ by itself. We take out of the scope that the factorization scale $\mu^2$ of hard processes would be chosen large enough, $\mu^2 \geq m_\rho^2$, where the $\rho$–meson mass $m_\rho$ represents the characteristic hadronic scale. Following this reason, the used coupling $\alpha_s(\mu^2)$ could not be too large.

## Table 1.

The results of $\Gamma_{(1,2)}(n)$ calculations ( $\Gamma(n) = \int_0^1 dx x^n P(z)$) performed in different ways, exact numerical results from [15] and approximation obtained from $P(z, A, \xi)$ with $\xi = -3$; both numerical and analytical **exact results** are emphasized by the bold print.

| | $\Gamma_{(1)}(n)$ | | $\Gamma_{(2)}(n)$ | | |
|---|---|---|---|---|---|
| | $C_F C_A$ | $N_f \cdot C_F$ | $C_A^2 C_F$ | $N_f \cdot C_F C_A$ | $N_f^2 \cdot C_F$ |
| n=2 **Exact** | **13.9** | **−2.3704** | **86.1 + 21.3 $\zeta(3)$** | **−12.9 − 21.3 $\zeta(3)$** | **−0.9218** |
| $\xi = -3$ | 11.3 | | −42.0 | 12.9 | |
| n=4: **Exact** | **23.9** | **−4.9152** | **140.0 + 19.2 $\zeta(3)$** | **−18.1 − 41.9 $\zeta(3)$** | **−1.5814** |
| $\xi = -3$ | 23.5 | | −76.0 | 23. | |
| n=6 **Exact** | **29.7** | **−6.4719** | **173 + 19.01 $\zeta(3)$** | **−20.4 − 54.0 $\zeta(3)$** | **−1.9279** |
| $\xi = -3$ | 31.1 | | −95.6 | 28.5 | |
| n=8 **Exact** | **33.9** | **−7.6094** | **196.9 + 18.98 $\zeta(3)$** | **−21.9 − 62.7 $\zeta(3)$** | **−2.1619** |
| $\xi = -3$ | 36.3 | | −109.0 | 32.3 | |
| n=10 **Exact** | **37.27** | **−8.5095** | **216.0 + 18.96 $\zeta(3)$** | **−23.2 − 69.6 $\zeta(3)$** | **−2.3366** |
| $\xi = -3$ | 41.00 | | −119.28 | 35.24 | |
| n=12 **Exact** | **40.02** | **−9.2555** | ? | ? | **−2.4753** |
| $\xi = -3$ | 44.64 | | −127.61 | 37.58 | |

To obtain the NNA result in a usual way, one should substitute the coefficient $b_0$ for $\gamma_g^{(q)}(0)$ into the expression for $A^{(q)}$ by hand (see, e.g., [10]). Note, the use of such an NNA procedure to improve

---
[5] Here, for the $\beta(a_s)$-function we adapt $\beta(a_s) = -b_0 a_s^2 + \ldots$, $b_0 = \frac{11}{3} C_A - \frac{4}{3} N_f T_R$

$P_q^{(1)}(z;A)$ leads to poor results even for $a_s^2 \, P_1(z)$ term of the expansion; a similar observation was also done in [17]. The NNA trick expresses common hope that the main logarithmic contribution may follow from the renormalization of the coupling constant. This renormalization appears as a sum of contributions from all the sources of renormalization of $a_s$, corresponding diagrammatic analysis for two-loop kernels is presented in [11, 4]. In the case of the $\xi = -3$ gauge the one-loop gluon renormalization imitates the contributions from these other sources and the coefficient $b_0$ appears naturally.

The expansion of kernel $P^{(1)}(z; A, \xi)$ generates partial kernels $a_s^2 P_{(1)}(z)$, $a_s^3 P_{(2)}(z), \ldots$ which in their turn produce ADCO $a_s^2 \, \Gamma_{(1)}(n)$, $a_s^3 \, \Gamma_{(2)}(n), \ldots$ according to the relation $\Gamma(n) = \int_0^1 dz z^n P(z)$. These elements of ADCO and a few numerical exact results from [16] are collected in Table 1, let us compare them:

(i) we consider there the contribution to the coefficient $\Gamma_{(1)}(n)$ which is generated by the gluon loops and associated with Casimirs $C_F C_A/2$, the $C_F^2$–term is missed, but its contribution is numerically insignificant. It is seen that in this order the $C_F C_A$–terms are rather close to exact values (the accuracy is about 10% for $n > 2$) and our approximation works rather well;

(ii) in the next order the contributions to $\Gamma_{(2)}(n)$ associated with the coefficients $N_f \cdot C_F C_A$ and $C_A^2 C_F$ are generated, while the terms with the coefficients $C_F^3$, $N_f \cdot C_F^2$, $C_F^2 C_A$ are missed. In the third order, contrary to the previous item, all the generated terms are opposite in sign to the exact values, and the "$\xi = -3$ approximation" doesn't work at all.

So, we need the next step to improve the agreement – to obtain the subleading $N_f$-terms by the **exact calculation**.

## 4  The non-forward ER-BL evolution kernel

Here we present the results of the bubble resummation for the ER-BL kernel $V(x, y)$. It can be derived in the same manner as it was done for the DGLAP kernel $P(z)$, see Appendix A in [3]. On the other hand $V(x, y)$ can be obtained as a "byproduct" of the previous results for $P(z)$, i.e., we shall use again [3] the exact relations between the $V$ and $P$ kernels established in any order of PT [11] for triangular diagrams. These relations were obtained by comparing counterterms for the same triangular diagrams considered in "forward", Fig.1a, and "nonforward", Fig.1d, kinematics.

Let the diagram in Fig.1a have a contribution to the DGLAP kernel in the form $P(z) = p(z) + \delta(1-z) \cdot C$; then its contribution to the ER-BL kernel (Fig.1d) is

$$V(x,y) = \mathcal{C}\left(\theta(y>x)\int_0^{\frac{x}{y}} \frac{p(z)}{\bar{z}} dz\right) + \delta(y-x)\cdot C, \tag{11}$$

where $\mathcal{C} \equiv 1 + (x \to \bar{x}, y \to \bar{y})$ to take into account the mirror-conjugate diagram. From relation (11) and Eqs. (3), (5) for $P^{(1a,c)}$ we immediately derive the expression for the sum of contributions $V^{(1a+1c)}$,

$$V^{(1a+1c)}(x,y;A,\xi) = a_s C_F 2 \cdot \mathcal{C}\left[\theta(y>x)(1-A)\left(\frac{x}{y}\right)^{1-A} - \frac{1}{2}\delta(y-x)\frac{(1-A)}{(2-A)}\right]\frac{\gamma_g(0,\xi)}{\gamma_g(A,\xi)}, \tag{12}$$

that may naturally be represented in the "plus form". Expression (12) can be independently verified by other relations reducing any $V$ to $P$ [11, 18] (see formulae for the $V \to P$ reduction there) and we came back to the same Eqs.(3), (5) for $P^{(1a,c)}$. Moreover, the first terms of the Taylor expansion of $V^{(1a,c)}(x,y;A)$ in $A$ coincide with the results of the two-loop calculation in

[11]. The relation $P \to V$ similar to Eq.(11) has also been derived for the diagram in Fig. 1b

$$V^{(1b)}(x,y) = \mathcal{C}\left[\theta(y > x)\frac{1}{2y}P^{(1b)}\left(\frac{x}{y}\right)\right]_+; \tag{13}$$

therefore, substituting Eq.(4) into (13) we obtain

$$V^{(1b)}(x,y;A,\xi) = a_s C_F 2 \cdot \mathcal{C}\left[\theta(y > x)\left(\frac{x}{y}\right)^{1-A}\frac{1}{y-x}\right]_+ \frac{\gamma_g(0,\xi)}{\gamma_g(A,\xi)}. \tag{14}$$

Collecting the results in (12) and (14) we arrive at the final expression for $V^{(1)}$ in the "main bubbles" approximation

$$V^{(1)}(x,y;A,\xi) = a_s C_F 2 \cdot \mathcal{C}\left[\theta(y > x)\left(\frac{x}{y}\right)^{1-A}\left(1 - A + \frac{1}{y-x}\right)\right]_+ \frac{\gamma_g(0,\xi)}{\gamma_g(A,\xi)}, \tag{15}$$

which has a "plus form" again due to the vector current conservation. The contribution $V^{(1)}$ in (15) should dominate for $N_f \gg 1$ in the kernel $V$. Besides, the function $V^{(1)}(x,y;A,\xi)$ possesses an important symmetry of its arguments $x$ and $y$. Indeed, the function $\mathcal{V}(x,y;A,\xi) = V^{(1)}(x,y;A,\xi) \cdot (\bar{y}y)^{1-A}$ is symmetrical under the change $x \leftrightarrow y$, $\mathcal{V}(x,y) = \mathcal{V}(y,x)$. This symmetry allows us to obtain the eigenfunctions $\psi_n(x)$ of the "reduced" evolution equation [4]

$$\int_0^1 V^{(1)}(x,y;A)\psi_n(y;A)dy = \Gamma(n;A)\psi_n(x;A), \tag{16}$$

$$\psi_n(y;A) \sim (\bar{y}y)^{d_\psi(A)-\frac{1}{2}} C_n^{d_\psi(A)}(y-\bar{y}), \text{ here } d_\psi(A) = (D_A - 1)/2, \quad D_A = 4 - 2A, \tag{17}$$

and $d_\psi(A)$ is the effective dimension of the quark field when the AD $A$ is taken into account; $C_n^{(\alpha)}(z)$ are the Gegenbauer polynomials of an order of $\alpha$. The partial solutions $\Phi(x;a_s,l)$ of the original ER-BL–equation ( where $l \equiv \ln(\mu^2/\mu_0^2)$)

$$\left(\mu^2\partial_{\mu^2} + \beta(a_s)\partial_{a_s}\right)\Phi(x;a_s,l) = \int_0^1 V^{(1)}(x,y;A)\,\Phi(y;a_s,l)dy \tag{18}$$

are proportional to these eigenfunctions $\psi_n(x;A)$ for the special case $\beta(a_s) = 0$, see, e.g. [3].

In the general case $\beta(a_s) \neq 0$ let us start with an ansatz for the partial solution of Eq.(18), $\Phi_n(x;a_s,l) \sim \chi_n(a_s,l) \cdot \psi_n(x;A)$, and the boundary condition is $\chi_n(a_s,0) = 1$; $\Phi_n(x;a_s,0) \sim \psi_n(x;A)$. For this ansatz Eq.(18) reduces to

$$\left(\mu^2\partial_{\mu^2} + \beta(a_s)\partial_{a_s}\right)\ln\left(\Phi_n(x;a_s,l)\right) = \Gamma(n;A). \tag{19}$$

In the case $n = 0$ the AD of the vector current $\Gamma(0;A) = 0$, and the solution of the homogeneous equation in (19) provides the "asymptotic wave function"

$$\Phi_0(x;a_s,l) = \psi_0(x;\bar{A}) \sim ((1-x)x)^{(1-\bar{A})}, \tag{20}$$

where $\bar{A} = -\bar{a}_s(\mu^2)\gamma(0,\xi)$ and $\bar{a}_s(\mu^2)$ is the running coupling corresponding to $\beta(a_s)$. A similar solution has been discussed in [10] in the framework of the standard NNA approximation. Solving simultaneously Eq. (19) and the renormalization-group equation for the coupling constant $\bar{a}_s$ we arrive at the partial solution $\Phi_n(x;\bar{a}_s,l)$ in the form

$$\Phi_n(x,\bar{a}_s) \sim \chi_n(\mu^2) \cdot \psi_n(x;\bar{A}); \text{ where } \chi_n(\mu^2) = \exp\left\{-\int_{a_s(\mu_0^2)}^{a_s(\mu^2)}\frac{\Gamma(n,A)}{\beta(a)}da\right\} \tag{21}$$

Recently, a form of the solution $\sim \psi_n(x;A)$ with $A = -a_s b_0$ has been confirmed in [12] by the consideration of conformal constraints [19] on the meson wave functions in the limit $N_f \gg 1$.

# 5 Conclusion

In this paper, I present closed expressions in the "all order" approximation for the DGLAP kernel $P(z)$ and ER-BL kernel $V(x,y)$ appearing as a result of the resummation of a certain class of QCD diagrams with the renormalon chain insertions. The contributions from these diagrams, $P^{(1)}(z;A)$ and $V^{(1)}(z;A)$, give the leading $N_f$ dependence of the kernels for a large number of flavours $N_f \gg 1$. These "improved" kernels are generating functions to obtain contributions to partial kernels like $a_s^{(n+1)} P_{(n)}(z)$ in any order $n$ of perturbation expansion. Here $A \sim a_s$ is a new expansion parameter that coincides (in magnitude) with the anomalous dimension of the gluon field. On the other hand, the method of calculation suggested in [3] does not depend on the nature of self-energy insertions and does not appeal to the value of the parameters $N_f T_R$, $C_A/2$ or $C_F$ associated with different loops. This allows us to obtain contributions from chains with different kinds of self-energy insertions, both quark and gluon (ghost) loops, see [8]. The prize for this generalization is gauge dependence of the final results for $P^{(1)}(z;A)$ and $V^{(1)}(z;A)$ on the gauge parameter $\xi$.

The result for the DGLAP non-singlet kernel $P^{(1)}(z;A(\xi),\xi)$ is presented in (6) in the covariant $\xi$-gauge. The analytic properties of this kernel in the variable $a_s$ are discussed for quark bubble chains only, and in the general case for an exceptional gauge parameter $\xi = -3$. For the latter case, $P^{(1)}(z;A(-3),-3)$ reproduces two-loop anomalous dimensions $a_s^2 \Gamma_{(1)}(n)$ with good accuracy, while the standard "naive nonabelianization" proposition fails at this level. But in the third order in $a_s$ the "$\xi = -3$ approximation" is insufficient, see quantities $\Gamma_{(2)}(n)$ in Table 1.

The contribution $V^{(1)}(x,y;A(\xi),\xi)$ to the non-forward ER-BL kernel (15) is obtained for the same classes of diagrams as a "byproduct" of the previous technique [8, 11]. The partial solutions (20, 21) to the ER-BL equation are derived.

The obtained results are certainly useful for an independent check of complicated computer calculations in higher orders of perturbation theory (PT), similar to [16]; they may be a starting point for further approximation procedures.

### Acknowledgements

The author is grateful to Dr. M. Beneke, Dr. K. Chetyrkin, Dr. A. Grozin, Dr. D. Mueller, Dr. A. Kataev, Dr. M. Polyakov, R. Ruskov and Dr.Dr. N. Stefanis for fruitful discussions of the results, author would like to thank the hosts of QUARKS'98 their extremely kind hospitality. This investigation has been supported by the Russian Foundation for Basic Research (RFBR) 98-02-16923.